\documentclass[twoside]{article}
\usepackage{fleqn,espcrc2}
\usepackage{graphicx}
\usepackage[figuresright]{rotating}

\newcommand{\AmS}{{\protect\the\textfont2
  A\kern-.1667em\lower.5ex\hbox{M}\kern-.125emS}}
\title{Angle-dependent magnetothermal conductivity in 
$d$-wave superconductors}

\author{I. Vekhter
	\address{Department of Physics, University of Guelph, Guelph,
			Ontario, Canada N1G 2W1}%
        \thanks{Acknowledges the hospitality
	of Centre \'Emile Borel and Aspen Center for Physics}
        and 
        P. J. Hirschfeld
	\address{Department of Physics, University of Florida,
        Gainesville, FL 32611-8440, USA}%
	\thanks{Supported in part by NSF Grant DMR-9974396}}
       
\begin{document}

\begin{abstract}
We analyse the behavior of the thermal conductivity, $\kappa(H)$,
 in the vortex state
of a quasi-two-dimensional d-wave superconductor when both the 
heat current and the applied magnetic field are in the basal
plane. At low temperature the effect of the field is accounted for 
in a semiclassical approximation, via a Doppler shift in the spectrum of the
nodal quasiparticles. In that regime $\kappa(H)$
exhibits twofold oscillations as a function of the angle between the 
direction of the field in the plane and the direction of the heat current, 
in agreement with experiment.
\vspace{1pc}
\end{abstract}
\maketitle

Experiments which show that the 
superconducting order
parameter in the high-T$_c$ cuprates 
is $d$-wave are often sensitive to the surface 
effects, and there is still
interest in the bulk probes of the 
symmetry of the gap. One piece of evidence for the linear
nodes in the bulk comes from the verification  of the universal
low temperature limit
of the in-plane thermal conductivity, \cite{louis}
another is based on the 
observed non-linear dependence of the electronic specific heat 
on the applied magnetic field, $H$, in the mixed state.\cite{moler}

The latter result is based on the observation that the properties
of a $d$-wave superconductor in the dilute vortex regime  ($H\ll H_{c2}$)
are determined by the near-nodal quasiparticles in the bulk.\cite{volovik}
In a semiclassical treatment, the energy of quasiparticle with the momentum
${\bf k}_n$, where $n$ labels a node, at point {\bf r} is shifted by
$\delta\omega={\bf v}_s({\bf r})\cdot {\bf k}_n$, where ${\bf v}_s$ is
the velocity field associated with the supercurrents. In the
regions
of the Fermi surface near the nodes where this shift exceeds the local gap,
there exist unpaired quasiparticles. Since the
typical supermomentum is $\hbar/R$ where
$R^2\approx \Phi_0/\pi H$, and since the number of 
vortices $n_v\propto H$, the spatially averaged density of states 
in a pure $d$-wave material varies as $\sqrt H$.
If the positions of the impurities and the vortices are uncorrelated,
the argument can be generalized to include the impurity scattering, 
which depends on the local density of states.\cite{kubert}
This simple approach has worked remarkably well in describing the 
low-temperature thermal and transport properties of the 
vortex state with the field perpendicular to the CuO$_2$ layers.
In particular, the increase of the $T=0$ limit of 
the in-plane thermal conductivity with $H$ is well described
by the semiclassical theory.\cite{kubert2} 

In relatively three-dimensional cuprates, such as YBa$_2$Cu$_3$O$_{7-\delta}$,
the same approach applies when the field is in the basal plane. In that
case the magnitude of the Doppler shift depends on the relative
orientation of the field with respect to the nodes, and the 
density of states in the pure limit exhibits fourfold oscillations as a
function of the angle between the direction of the field and the crystalline
axes.\cite{vekhter} Experimental verification of this prediction 
has not been possible so far and is
hindered by the 
extrinsic contributions to the specific heat\cite{junod} and 
the orthorhombicity of the material.\cite{carbotte}
On the other hand, the angular dependence of the thermal conductivity
in the vortex state has already been observed,
\cite{yu,aubin} 
and here we analyse it
in the semiclassical framework.

As in Refs.\cite{kubert2,vekhter} we assume a
cylindrical Fermi surface and a $d$-wave gap,
 and approximate the superflow
by the velocity field around a single vortex. 
The field and the thermal gradient
are applied at  angle $\alpha$ and $\varepsilon$ to the $\widehat{\rm a}$ axis
respectively.
We consider the regime  $T,\gamma\ll E_H \ll\Delta_0$, where
$E_H\approx (v_f/2)(\pi\Phi_0\lambda_{ab}/H\lambda_c)^{1/2}$ is
the average Doppler shift, and $\gamma$ is the low-energy scattering rate.
Following Refs.\cite{kubert2,vekhter}
we obtain that the local change in $\kappa$
is
$\delta\kappa (\rho)/\kappa_{00}
=E(\alpha)\sin^2\beta/\rho^2$, where
$\kappa_{00}/T= \pi N_0 v_f^2/6\Delta_0$ 
is the universal thermal conductivity,
$N_0$ is the normal state density of states, $\Delta_0$ is the gap amplitude,
$v_f$ is the Fermi velocity,
$\beta$ is the winding angle of the vortex,
$\rho$ is the distance from the center of the vortex normalized to
$R$,
$E(\alpha)=
(\pi E_H^2/ 8\Gamma\Delta_0)
\max(\sin^2\alpha, \cos^2\alpha)$,
and $\Gamma$ is the bare scattering rate.
The local $\kappa({\bf r})$ has to be spatially averaged to obtain the field
dependence. When the heat gradient, $\nabla T$,
 is parallel to
the field $\kappa_{\|}(H)=\langle \kappa({\bf r})\rangle$, where the brackets 
denote the average
over a unit cell of the vortex lattice.
 For other relative orientations of $\nabla T$ and 
{\bf H} the averaging procedure is not clear; it
was argued\cite{kubert2} that 
$\kappa_\perp(H)=[\langle (1/\kappa({\bf r}))\rangle ]^{-1}$
is appropriate for
$\nabla T\perp {\bf H}$.
We therefore take here a simple approach of averaging 
independently the components of the heat current along and normal to the
vortex, and expect that this procedure gives at least
qualitatively correct
results. Then the longitudinal and the Hall thermal conductivity are
given by $\kappa_1=\kappa_{\|}\cos^2(\alpha-\varepsilon)
+\kappa_\perp \sin^2(\alpha-\varepsilon)$ and
$\kappa_2=(1/2)|(\kappa_\perp-\kappa_{\|})\sin 2(\alpha-\varepsilon)|$
respectively, with
$\kappa_{\|}=\kappa_{00}[1+E(\alpha)\ln(\Delta_0/E_H)]$ and
\begin{equation}
{\kappa_{\perp}\over\kappa_{00}}=\biggl[\sqrt{1+E(\alpha)}-
		E(\alpha)\sinh^{-1}{1\over\sqrt{E(\alpha)}}\biggr]^{-1}.
\end{equation}
The result for $\kappa_1$ is in agreement with the twofold pattern
of Ref.\cite{aubin} at $T=0.8$K with
$\varepsilon=\pi/2$, see Fig.1.
The minima of $\kappa_1(\alpha)$ correspond to increased
scattering by the vortices when the heat current is normal to the field.

In the regime where $T\ge E_H$
Yu {\it et al.} \cite{yu}
have measured $\kappa_\pm=(\kappa_1\pm\kappa_2)/\sqrt 2$
with $\varepsilon=0$,
found a twofold pattern,
and explained it as a consequence of Andreev reflection
of quasiparticles in the presence of supercurrents;
we note that the angular dependence of $\kappa_\pm$
is very similar to what we obtain at $E_H\gg T$.
In the same regime
Aubin {\it et al.} \cite{aubin}
observed a {\it fourfold} symmetry of $\kappa_1(\alpha)$,
consistent with the picture in which the 
scattering of quasiparticle by the vortices becomes more important at
high $T$\cite{vekhter2} 
and this scattering has the same symmetry as the gap.\cite{yu,vekhter2}
\begin{figure}
\includegraphics[width=7cm]{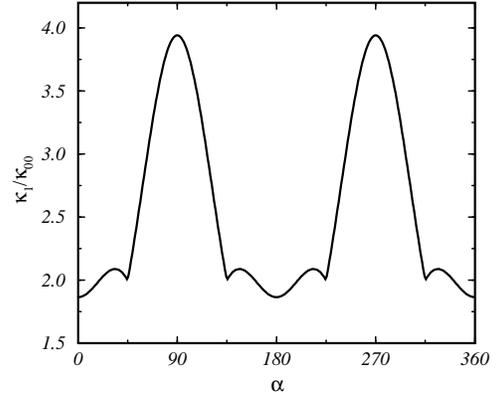}
\caption{Angular dependence of $\kappa_1$ for
$E_H=0.05\Delta_0$ and $\Gamma=0.001\Delta_0$.}
\end{figure}
The work to explain in detail the high $T$ dependence is in progress, 
and will be reported elsewhere.

\end{document}